\documentclass[aps,prd,twocolumn,superscriptaddress,showpacs,floatfix,nofootinbib]{revtex4-1}
\usepackage{amsmath,graphicx}
\usepackage{hyperref}
\usepackage{orcidlink}
\usepackage[x11names]{xcolor}
\usepackage{bm}
\usepackage{amsfonts,amssymb}
\usepackage{subcaption} 

\def\TRENTo{{\sc t\kern-.05em \lower.5ex\hbox{r}\kern-.025em e\kern-.05em n\kern-.05em t\kern-.09em}o}
\def\iccing{{\sc i\kern-.05em c\kern-.05em c\kern-.05em i\kern-.05em n\kern-.05em g\kern-.05em}}
\def\ccake{{\sc c\kern-.05em c\kern-.05em a\kern-.05em k\kern-.05em e\kern-.05em}}
\def\vUSPhydro{v-{\sc u\kern-.05em s\kern-.05em p\kern-.05em}hydro}
\graphicspath{ {figures/} }

\begin{document}

\title{Symmetry Energy of 2+1-flavor dense quark matter from perturbative QCD}

\author{Isabella Danhoni \, \orcidlink{0000-0003-0126-393X}}
\email{idanhoni@illinois.edu }
\author{Yumu 
Yang\,\orcidlink{0009-0001-8979-9343}}
\email{yumuy2@illinois.edu}
\affiliation{The Grainger College of Engineering, Illinois Center for Advanced Studies of the Universe, Department of Physics, University of Illinois at Urbana-Champaign, Urbana, IL 61801, USA}
\author{Mauricio Hippert\,\orcidlink{0000-0001-5802-3908}}
\email{hippert@cbpf.br}
\affiliation{Centro Brasileiro de Pesquisas Físicas, Rua Dr. Xavier Sigaud 150, 
Rio de Janeiro, RJ, 22290-180, Brazil}
\author{Jacquelyn Noronha-Hostler\,\orcidlink{0000-0003-3229-4958}}
\email{jnorhos@illinois.edu}
\affiliation{The Grainger College of Engineering, Illinois Center for Advanced Studies of the Universe, Department of Physics, University of Illinois at Urbana-Champaign, Urbana, IL 61801, USA}

\date{\today}

\begin{abstract}
The symmetry energy expansion was developed to connect isospin symmetric matter probed in nuclear experiments to asymmetric matter found in neutron stars. Using the isospin asymmetry derived from the Gell-Mann-Nishijima formula, we derive the symmetry energy expansion for quark matter that has unique properties compared to hadronic matter. To test our methods, we use perturbative Quantum Chromodynamics (pQCD) calculations at next-to-leading-order, where realistic quark masses can be included. We find that pQCD at electroweak equilibrium is not isospin symmetric but rather obtains a small skewness term in the symmetry energy expansion. We predict that if equations of state for nuclear matter must match pQCD results, then a non-monotonic dip in the symmetry energy would appear. 
\end{abstract}

\maketitle


\section{ Introduction}

The dense matter equation of state (EOS) can be probed both in the laboratory using nuclear experiments or in astrophysical phenomena such as neutron stars, binary neutron star mergers, or supernovae. While experiments and astrophysical phenomena reach similar baryon densities $n_B$, they probe very different regimes of the Quantum Chromodynamics (QCD) phase diagram because of isospin. Nuclei are approximately isospin symmetric i.e. they contain a nearly equal number of protons and neutrons. Given that protons contain $uud$ quarks and neutrons contain $ddu$ quarks, the isospin symmetry is also held at the quark level, since there is nearly an equal number of up and down quarks. However, neutron stars undergo inverse $\beta$ decays early in their evolution, such that they become neutron rich, leading to a strong isospin asymmetry. At the quark level within neutron stars this picture is significantly more complicated, which will be explored here.

To connect isospin symmetric matter to isospin asymmetric matter, historically the symmetry energy expansion \cite{Bombaci:1991zz} has been a powerful tool. 
The symmetry energy expansion expands the binding energy per baryon $E/A$ around symmetric nuclear matter (SNM) out to some arbitrary isospin asymmetry $\delta_I$ to obtain asymmetric nuclear matter (ANM). 
Typically this expansion is taken up to $\delta_I^2$
\begin{equation}\label{eqn:symexpan}
    \frac{E_{ANM}}{A}(\delta_I)=\frac{E_{SNM}}{A}+\tilde{E}_{sym,2}\delta_I^2+\tilde{E}_{sym,3}\delta_I^3 +\mathcal{O}(\delta_I^4)
\end{equation}
where the linear term drops as long as one correctly defines the isospin asymmetry, and it is understood that all terms depend on $n_B$, which is not shown for simplicity's sake. 
Here the symmetry energy $\tilde{E}_{sym,2}(n_B)$ encodes the difference between SNM and ANM. 
The symmetry energy has very successfully described proton and neutron matter across a range of $\delta_I$'s \cite{Tsang:2008fd,Kortelainen:2010hv,Kortelainen:2011ft,Russotto:2011hq,Brown:2013mga,Holt:2013fwa,Danielewicz:2013upa,Baldo:2016jhp,Zhang:2015ava,Holt:2016pjb,Zhang:2018vrx,Drischler:2020yad,Drischler:2021kxf,Lynch:2021xkq,Yao:2023yda,Sorensen:2023zkk,Lattimer:2023rpe,Brown:2024rml,Yang:2025wop}, even when only keeping terms up to $\delta_I^2$. 
From this point onward, we loosely use the terms {\bf symmetric nuclear matter} (SNM) $\delta_I=0$, {\bf asymmetric nuclear matter} (ANM) $\delta_I>0$, and {\bf pure nuclear matter} (PNM) $\delta_I=1$ throughout this paper since they are commonly used acronyms within the community, even though we will be referring to dense hadronic matter (that may include hyperons) and dense quark matter as well.

However, the original symmetry energy expansion assumed only nuclear matter where the presence of strange particles was not necessary to describe isospin asymmetry.  In quark matter, strange quarks are abundant so we must rely on the Gell-Mann-Nishijima formula to derive the isospin asymmetry term, as was done for the first time in \cite{Yang:2025wop} for hadronic matter. In this approach, the isospin asymmetry is described by:
\begin{equation}
    \delta_I=1-2Y_Q+Y_S
\end{equation}
where the electric charge fraction $Y_Q=n_Q/n_B$ and the strangeness fraction  $Y_S=n_S/n_B$ are normalized by the baryon number density. 
In the limit where the strangeness content is in electroweak equilibrium, a skewness term appears in the symmetry energy expansion (i.e. $\tilde{E}_{sym,3}(n_B)\neq 0$) due to the presence of $Y_S\neq 0$. 

Previous attempts have been made to determine the symmetry energy of quark matter (or its influence on the presence of quark matter) using the original symmetry energy expansion \cite{Kim:2010dp,Chu:2012rd,Jeong:2015ima,Chen:2017och,Sahoo:2021fxw,Stone:2022unw,Kumar:2023qcs,Ju:2025hqm,Divaris:2025bsz}. In systems that only contain up and down quarks, these would be entirely consistent with the new approach in \cite{Yang:2025wop} to describe isospin asymmetry. However, in the presence of strangeness (either hyperons or strange quarks), one must use the correct formalism found in \cite{Yang:2025wop} --- otherwise one does not even obtain the ground state of matter for isospin symmetric matter. In these previous works, a symmetry energy of quarks was extracted to be in the order of hundreds of MeV's.

Meanwhile, perturbative QCD (pQCD) \cite{Freedman:1976ub,Freedman:1976xs,Kurkela:2009gj} has generated much interest due to the possibility of very high-order calculations \cite{Gorda:2018gpy,Gorda:2021gha,Gorda:2021kme,Gorda:2023mkk,Karkkainen:2025nkz} that appear to converge \cite{Gorda:2023mkk}. 
Using causality and stability constraints \cite{Komoltsev:2021jzg}, they have been applied to attempt to constrain the equation of state (EOS) of neutron stars at much lower $n_B$ \cite{Gorda:2022jvk,Somasundaram:2022ztm,Mroczek:2023zxo,Komoltsev:2023zor}. 
However, the N3LO (next-to-next-to-next-to-leading-order) calculations were performed in the limit of massless quarks where the quark number densities become identical i.e. $n_u=n_d=n_s$ in $\beta$-equilibrium. In this limit, there is no isospin asymmetry, since $\delta_I=0$.

In reality, the flavor symmetry between up ($u$), down ($d$), and strange ($s$) quarks is broken by finite quark masses. Because strange quarks are heavier, their mass plays the largest role. Intuitively, we can understand this because a massive strange quark with $m_s\neq 0$ will suppress the number density of strange quarks such that $n_u>n_s$ and $n_d>n_s$. Even the light quarks masses could play a small role, such that one expects a hierarchy (at least from their vacuum masses) where $n_u>n_d>n_s$. Unfortunately, pQCD calculations that allow for finite quark masses do not yet exist at N3LO but there are NLO calculations available from \cite{Graf:2015tda} for one massive quark.

In this paper we use pQCD results up to next-to-leading order for $T=0$~\cite{Graf:2015tda} to lay out the framework for symmetry energy calculations of cold, dense quark matter. 
 We confirm previous results from strange hadronic phases \cite{Yang:2025wop} that the presence of strangeness leads to a skewness term in the symmetry energy expansion in the case of weak-equilibrium also for quark phases. 

\section{pQCD at $T=0$ across $\mu_B,\mu_S,\mu_Q$}
\label{sec:pqcd_definitions}

Here we review the results obtained in~\cite{Graf:2015tda} for the NLO thermodynamical potential derived for pQCD. As noted in their work, the zero-temperature $T=0$ case can be computed analytically for 2+1 flavors, such that one can directly use their pocket formula for the thermodynamical potential to compute the limits for pressure $p$ and energy density $\varepsilon$ in pQCD. Since this expression considers only one massive quark, here we focus on the effects of the strange-quark mass for the symmetry energy. We reproduce this expression below,
\begin{align}
  \label{eq-ThermPotTZero}
  \begin{split}
    \Omega^{(0)}_f=&\frac{-N_c}{12\pi^2}\left[\mu_f k_f\left(\mu_f^2-\frac{5}{2}m_f^2\right)+\frac{3}{2}m_f^4\ln\left(\frac{\mu_f+k_f}{m_f}\right)\right], \\
    \Omega^{(1)}_f=&\frac{\alpha_sN_G}{16\pi^3}\left\{3\left[m_f^2\ln\left(\frac{\mu_f+k_f}{m_f}\right)-\mu_f k_f\right]^2-2k_f^4\right. \\
    +&\left.m_f^2\left[6\ln\left(\frac{\Lambda}{m_f}\right)+4\right]\left[\mu_f k_f-m_f^2\ln\left(\frac{\mu_f+k_f}{m_f}\right)\right]\right\},
  \end{split}
\end{align}
%
where $N_c$ is the color degeneracy, $N_G = N_c^2 - 1$ is the number of gluons, $\Lambda$ is the renormalization scale, $\Omega^{(0)}$ is the contribution from the free quarks (or the free gas term),  $\Omega^{(1)}$ is the result of the integration over the renormalized exchange contribution (the NLO contribution), and, finally, $k_f\equiv\sqrt{\mu_f^2-m_f^2}$, for each flavor $f$ . The running coupling is incorporated based on the prescription from Ref.~\cite{Fraga:2004gz}, where $\alpha_s(\Lambda)$ is given by,
\begin{equation}
\alpha_s(\Lambda)=\frac{4\pi}{\beta_0\bar{L}}\left[1-2\frac{\beta_1}{\beta_0^2}\frac{\ln \bar{L}}{\bar{L}}\right],
\label{eq-RunAlphas}
\end{equation}
which allows one to compute the strange-quark mass as a function of the strong coupling as,
\begin{equation}
m_s(\Lambda)=\hat{m}_s\left(\frac{\alpha_s}{\pi}\right)^{4/9}\left[1+0.895062\frac{\alpha_s}{\pi}\right],
\label{eq-SMass}
\end{equation}
where $\bar{L}=2\ln(\Lambda/\Lambda_{\overline{\text{MS}}})$, $\beta_0=11-2N_f/3$ and $\beta_1=51-19N_f/3$. The scale $\Lambda_{\overline{\text{MS}}}$ and the invariant mass $\hat{m}_s$ are fixed according to the same definitions used in~\cite{Graf:2015tda}, so that our calculation of the pressure agrees with what was obtained in their work.%
\footnote{%
We have tested our results against figures that appeared originally in \cite{Graf:2015tda} and confirmed that we can reproduce their results. }

Since in this work we consider large baryon densities, the scale is essentially set by the baryon chemical potential $\mu_B$. 
However, because we are interested in the effects of flavor-symmetry breaking due to quark masses, we must be careful to choose a renormalization scale which is invariant under the exchange of the $u$ and $d$ quarks, similarlly to what was done in~\cite{Andersen:2015eoa}. 
We thus use the baryon chemical potential at fixed strangeness $S$ and isospin along the up-down direction $I_z$,
\begin{equation}
\tilde\mu_B^{(I_z,S)}\equiv \left(\frac{\partial \epsilon}{\partial n_B}\right)_{n_I,n_S} = \mu_B + \frac{\mu_Q}{2},
\end{equation}
where $\epsilon$ is the energy density. 
Unlike $\mu_B\equiv (\partial \varepsilon/\partial n_B)_{n_Q,n_S}$, the combination $\tilde\mu_B= \mu_B + \mu_Q/2$ is invariant under an isospin reversal operation given by $I_z\to-I_z$, $B\to B$, $S\to S$, 
where $B$ is the baryon number (see the discussion and appendix in \cite{Yang:2025wop}).
Therefore, we choose the renormalization scale to be 
\begin{equation}
\Lambda=2\sqrt{(\pi T)^2+((\mu_B+\mu_Q/2)/3)^2}.
\label{Eq:lambda}
\end{equation}

The pressure is given by
\begin{equation}
\begin{aligned}
    P =-\Omega^{tot}\equiv
    - \sum_{f=u,d,s} \left(
 \Omega_f^{(0)}(m_f, \mu_f) 
+ \Omega_f^{(1)}(m_f, \mu_f) \right)
\label{eq:pressure}
\end{aligned}
\end{equation}
where the sum is over different flavors $f$.
In order to ensure thermodynamical consistency, 
we ensure that
\begin{equation}
    n_f=\frac{\partial P}{\partial \mu_f}\Bigg |_{T\,,\,n_{f\neq{f'}}}\,, 
\end{equation}
including indirect contributions from the derivatives of $\Lambda(\mu_B,\mu_Q)$,    
and obtain the energy density from the Gibbs-Duhem relation at vanishing $T$:
\begin{eqnarray}
\epsilon=n_B\mu_B+n_S\mu_S+n_Q\mu_Q-{P},
\end{eqnarray}
or, for the extraction of the symmetry energy,
\begin{eqnarray}
    \frac{\epsilon}{n_B}=\mu_B+Y_S\mu_S+Y_Q\mu_Q-\frac{{P}}{n_B}.
\end{eqnarray}

In the case of electroweak equilibrium our energy density simplifies to: $\epsilon=n_B\mu_B+n_Q\mu_Q-\bar{P}$.



\section{Isospin asymmetry of quarks}

The N2LO pQCD calculations of cold, dense matter in  \cite{Komoltsev:2021jzg} were taken in the limit where the three relevant (massless) quarks:  $u$, $d$ and $s$  have exactly equal populations i.e. $n_u=n_d=n_s$, which arises from setting their chemical potentials equal $\mu_u=\mu_d=\mu_s=\mu_B/3$. 
To understand this assumption better, we can use a massless ideal Fermi gas as an example.  The number density of a massless ideal Fermi gas is given by:
\begin{equation}
    n_i=\frac{1}{3\pi^2}\mu_i^3
\end{equation}
such that setting the same chemical potential for each flavor is equivalent to setting the same number density (note this holds also if all three quarks have the same mass but breaks for different masses). 
In this limit, the system is exactly electrically neutral since 
\begin{equation}
    n_Q=+\frac{2}{3}n_u-\frac{1}{3}\left[n_d+n_s\right]
\end{equation}
such that when $n_u=n_d=n_s$, then $n_Q=0$, $Y_Q=0$, and $\delta_I=0$. For massless quarks, the concept of $\beta$-equilibrium does not exist because no leptons are required in the system in order to obtain electric charge neutrality.

Here we relax the assumption of three massless quarks to explore the symmetry energy of pQCD.  
We note that the symmetry energy cannot be defined in the typical manner as what was originally done in \cite{Bombaci:1991zz}. 
The work in \cite{Chu:2012rd} get closest to what we will do here but we note that their definition of isospin asymmetry in Eq.\ (3) of their work ignores the baryon contribution of strange quarks. Additionally, they did not explore the possibility of a skewness term in their symmetry energy expansion.

The original symmetry energy expansion focused on nuclei such that the degrees of freedom were limited to only protons and neutrons, which simplified certain limits. 
For instance, one could define the isospin asymmetry coefficient of a system that only includes protons and neutrons as
\begin{equation}
    \delta_Q=1-2Y_Q.
\end{equation}
For such a system, SNM corresponds to the limit $\delta_Q = 0$, and PNM corresponds to the limit $\delta_Q = 1$. 
In terms of $u$ and $d$ quark flavors, these limits correspond to $n_u=n_d$ and $2n_d=n_u$, respectively. 

\begin{figure}
    \centering
    \includegraphics[width=\linewidth]{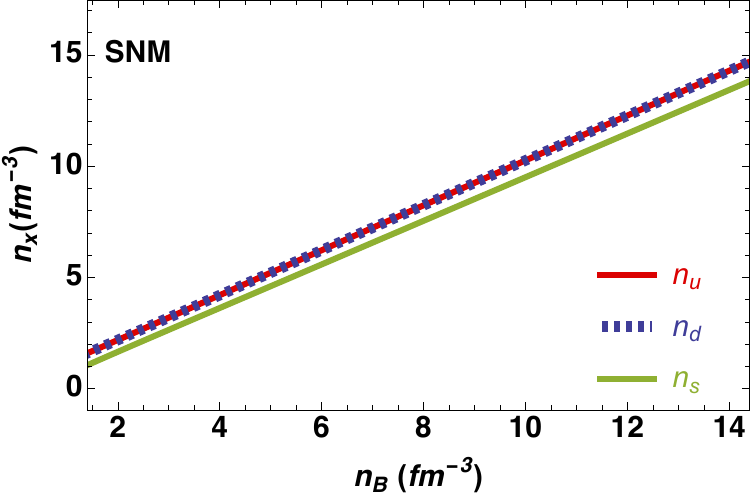}
    \caption{Population plots of up, down, strange quarks for SNM $\delta_I=0$ vs baryon number density. 
    }
    \label{fig:SNM_pops}
\end{figure}

\begin{figure}
    \centering
    \includegraphics[width=\linewidth]{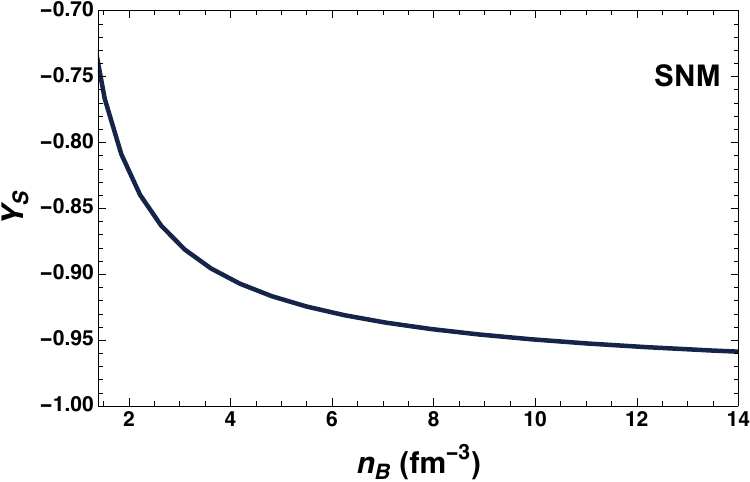}\\
    \includegraphics[width=\linewidth]{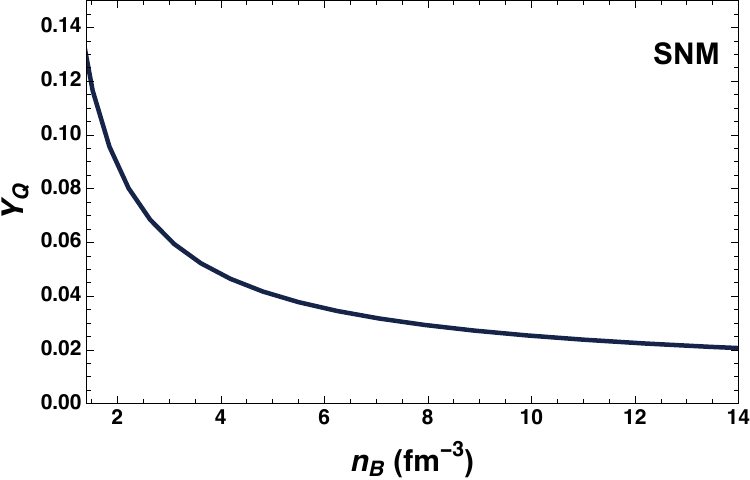}
    \caption{
 (Top) Strangeness to baryon number $Y_S(n_B)$ and (bottom) electric charge to baryon number density $Y_Q(n_B)$  vs baryon number density, for symmetric nuclear matter, i.e. $\delta_I=0$. Calculated in NLO pQCD. 
    }
    \label{fig:SNM_YSnb}
\end{figure}


However, in the presence of strangeness \cite{Yang:2025wop},  isospin asymmetry must incorporate both electric charge $Q$ and strangeness $S$, such that the isospin asymmetry coefficient is defined as:
\begin{equation}
\label{eq:deltaI}
    \delta_I=1-2Y_Q+Y_S.
\end{equation}
In term of quark densities we can rewrite the fractions as:
\begin{eqnarray}
    Y_Q&=&\frac{\frac{2}{3}n_u-\frac{1}{3}n_d-\frac{1}{3}n_s}{\frac{1}{3}n_u+\frac{1}{3}n_d+\frac{1}{3}n_s}=\frac{2n_u-n_d-n_s}{n_u+n_d+n_s}\\
    Y_S&=&\frac{-n_s}{\frac{1}{3}n_u+\frac{1}{3}n_d+\frac{1}{3}n_s}=-3\frac{n_s}{n_u+n_d+n_s}
\end{eqnarray}
such that the isospin asymmetry becomes:
\begin{eqnarray}
    \delta_I&=&3\frac{n_d-n_u}{n_u+n_d+n_s}.
\end{eqnarray}
where the factor of 3 appears due to the weight of the baryon charge $B_f=1/3$ that each quark carries. We note that \cite{Chu:2012rd} defines isospin asymmetry in their Eq. (3) the closest to what we have in this work. However, they chose to separate $n_s$ in their expansion instead of including it into $\delta_I$ such that their normalization is not actually baryon density but only that of light quarks. 

\subsection*{ SNM of 2+1 flavor weakly interacting quarks}
Once one redefines the isospin asymmetry coefficient and $Y_S\neq 0$, previous limits no longer hold. 
SNM can still be obtained for $\delta_I=0$,  but $Y_Q^{SNM}\neq 0.5$. In fact, since $Y_S<0$ then $Y_Q^{SNM}=0.5+0.5\,Y_S^{SNM}$ such that $Y_Q^{SNM}<0.5$ for finite strangeness.  
In terms of quarks, SNM for $\delta_I=0$ implies that $n_u=n_d$ but there are no constraints on $n_s$ such that 
\begin{equation}
    Y_Q^{SNM}=\frac{n_u-n_s}{2n_u+n_s}\\
\end{equation}
where it should be clear that the larger values of $n_s$ for SNM leads to smaller values of $Y_Q^{SNM}$.

In Fig.\ \ref{fig:SNM_pops} we confirm the above in pQCD for SNM under electroweak equilibrium for the strangeness content, i.e. $\mu_S=0$. That is, we see that when $\delta_I=0$ we obtain exactly identical populations of $n_u=n_d$ as a function of $n_B$. We note that with increasing $n_B$ the value of $n_s$ increases essentially linearly.
In Fig.\ \ref{fig:SNM_YSnb}, we can also observe fractions of conserved charges to baryon number density i.e. $Y_S^{SNM}(n_B)$ (top) and $Y_Q^{SNM}(n_B)$ (bottom) for SNM, again under strange electroweak equilibrium. 
For SNM we have the constraint of:
\begin{equation}
    1+Y_S^{SNM}=2\,Y_Q^{SNM}
\end{equation}
such that, the more negative $Y_S$, the smaller $Y_Q$, with $Y_S^{\textrm{conf}}\rightarrow -1$ leading to $Y_Q^{\textrm{conf}}\rightarrow 0$ in the massless (or conformal) limit.
On the other hand, when the strange quark mass plays a larger role, we expect $Y_Q$ to increase. 
In Fig.\ \ref{fig:SNM_YSnb} we can clearly see that the strangeness fraction approaches the conformal limit of  $Y_S\rightarrow -1$ at high $n_B$.  
Similarly, for the electric charge fraction we see precisely the behavior that we expect in that it approaches zero at large $n_B$ but $Y_Q$ becomes larger at lower $n_B$. 
However, even at very high $n_B=12 \text{ fm}^{-3}\sim 75 \; n_{sat}$, we find a small deviation from this limit. 
We note that the population of strange quarks consistently remains below that of up and down quarks throughout, even though the relative difference decreases with $n_B$. 

\begin{figure}
    \centering
    \includegraphics[width=\linewidth]{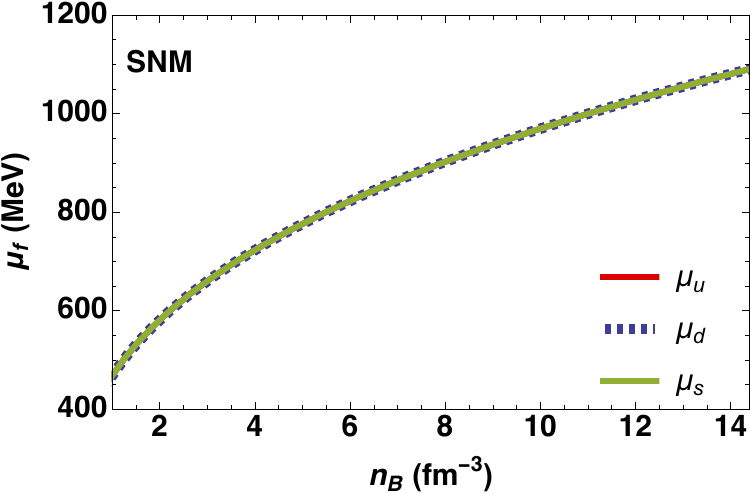}
    \caption{Quark chemical potentials for up quarks $\mu_u$, down quarks $\mu_d$, and strange quarks $\mu_S$ vs baryon number density for SNM. }
    \label{fig:SNM_chemical_potentials}
\end{figure}

In Fig.\  \ref{fig:SNM_chemical_potentials} we plot the relationship between the quark chemical potentials $\mu_f$ to that of $n_B$, again for SNM. As one would expect, all the quark chemical potentials are identical in this limit i.e. $\mu_u=\mu_d=\mu_s$, which works out to $\mu_S=\mu_Q=\mu_I=0$. 

\begin{figure}
    \centering
    \includegraphics[width=\linewidth]{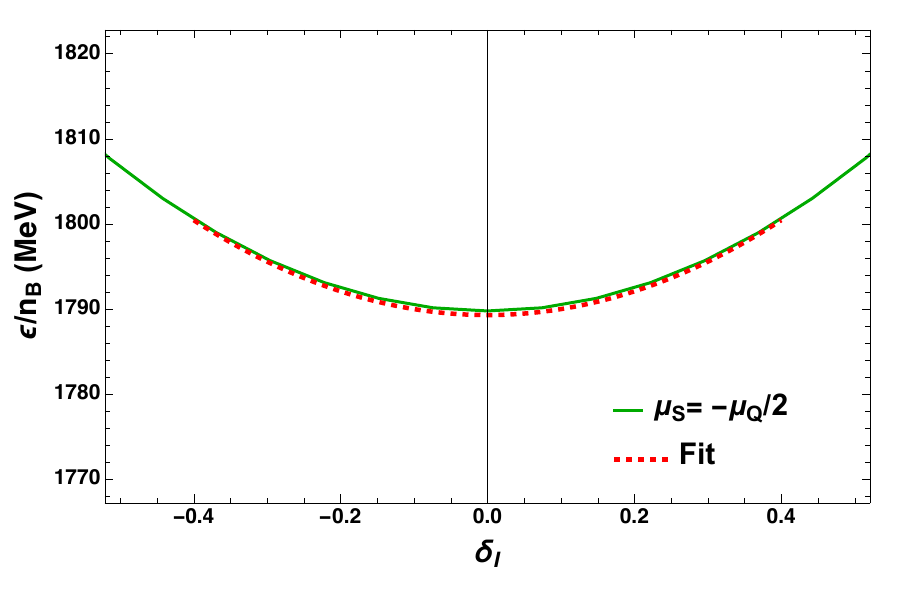}
    \caption{E/nB vs delta for some fit
    }
    \label{fig:E_nb}
\end{figure}


\subsection*{Range of isospin asymmetry in a 2+1 flavor quark model}
The other limit of $\delta_Q=1$ would imply PNM only for a system with just up and down quarks where $2n_d=n_u$ and $n_s=0$, such that $Y_Q=0$. 
Otherwise, all that $\delta_I=1$ implies is that:
\begin{equation}
    n_s=2n_d-4n_u,
\end{equation}
which is not a particularly useful limit. 
Rather, it is more interesting to think about the maximum isospin asymmetry $\delta_I^{max}$ that appears when only down quarks are present:
\begin{eqnarray}
    \delta_I^{max}&=&3.
\end{eqnarray}
Conversely, we can also obtain our most negative $\delta_I$ as well when we have only up quarks, $\delta_I\rightarrow -3$.
The factor of $\pm3$ appears because of the $1/3$ baryon number that each quark carries.

\section{Charge and Strangeness fraction phase diagram}

\begin{figure*}
\begin{tabular}{cc}

      \includegraphics[width=0.53\linewidth]{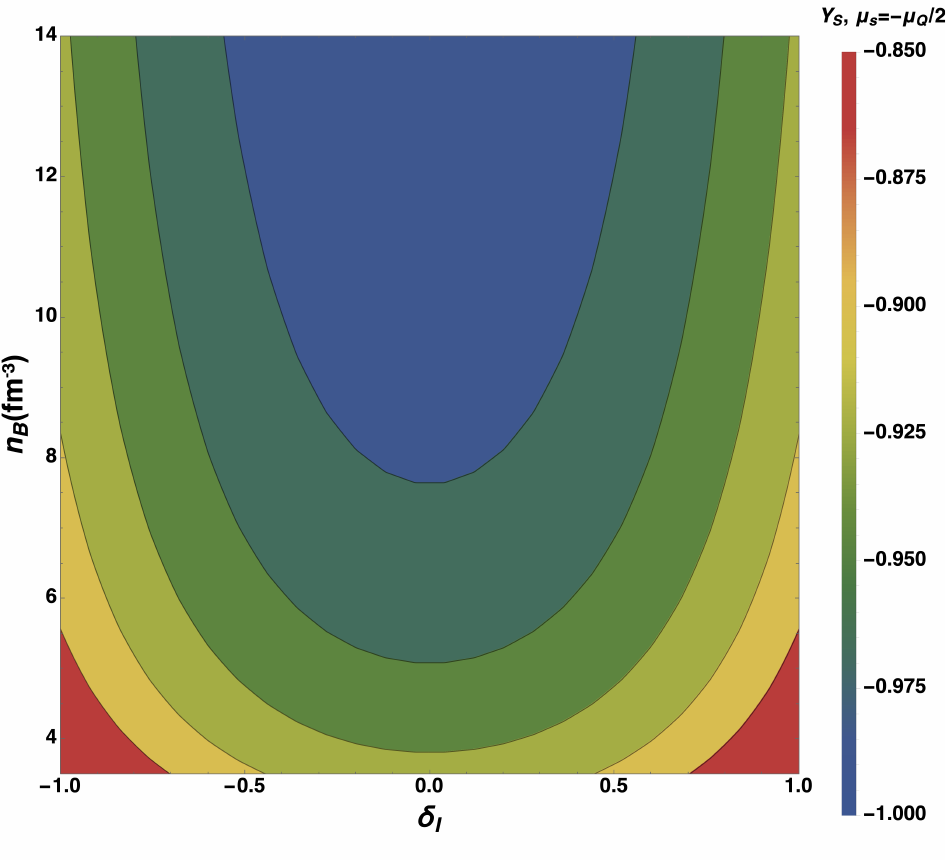} &\includegraphics[width=0.5\linewidth]{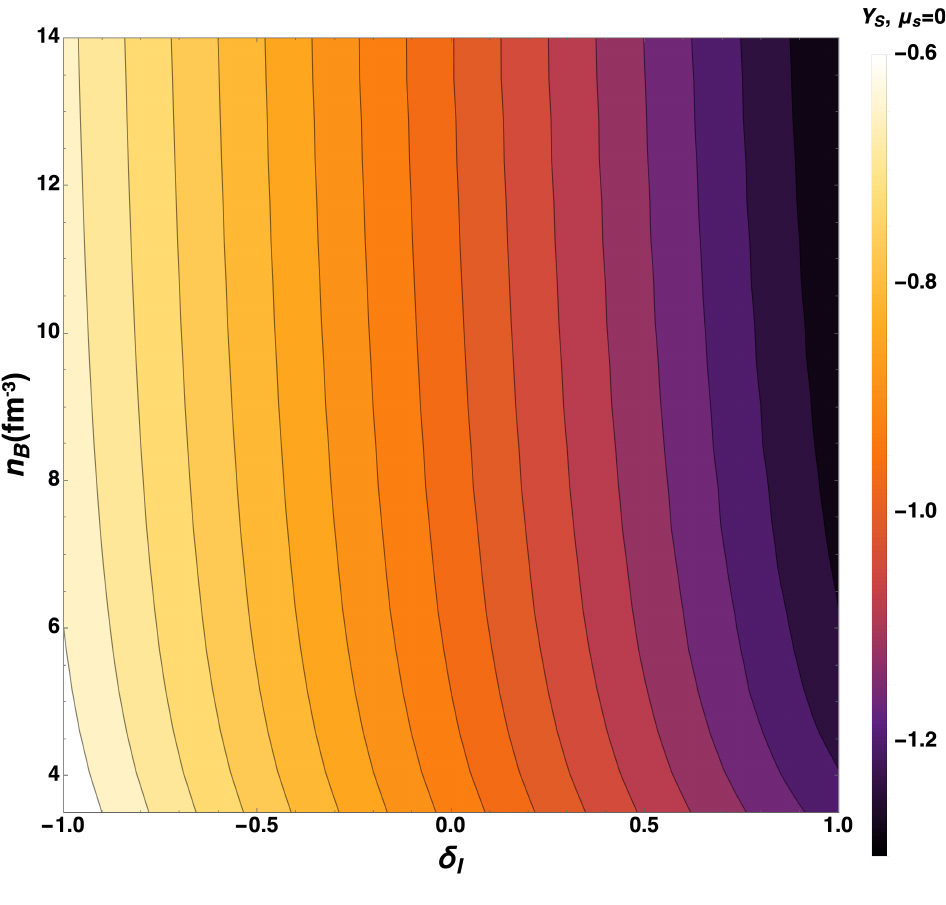}     \\     \includegraphics[width=0.53\linewidth]{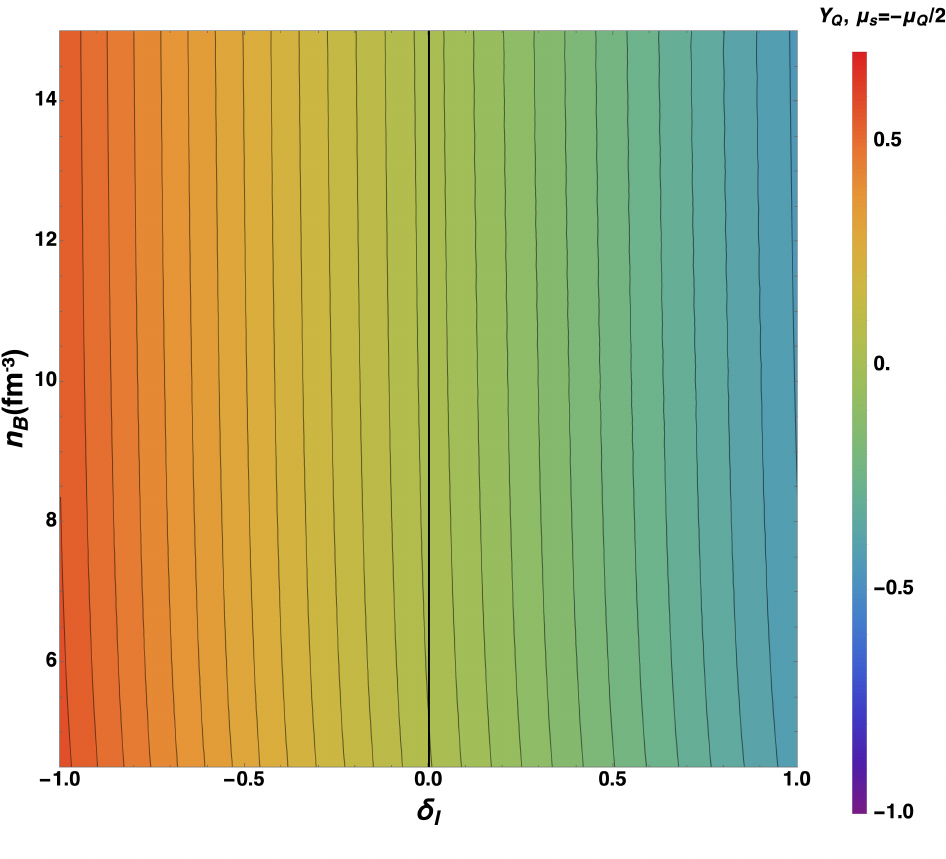} &  \includegraphics[width=0.5\linewidth]{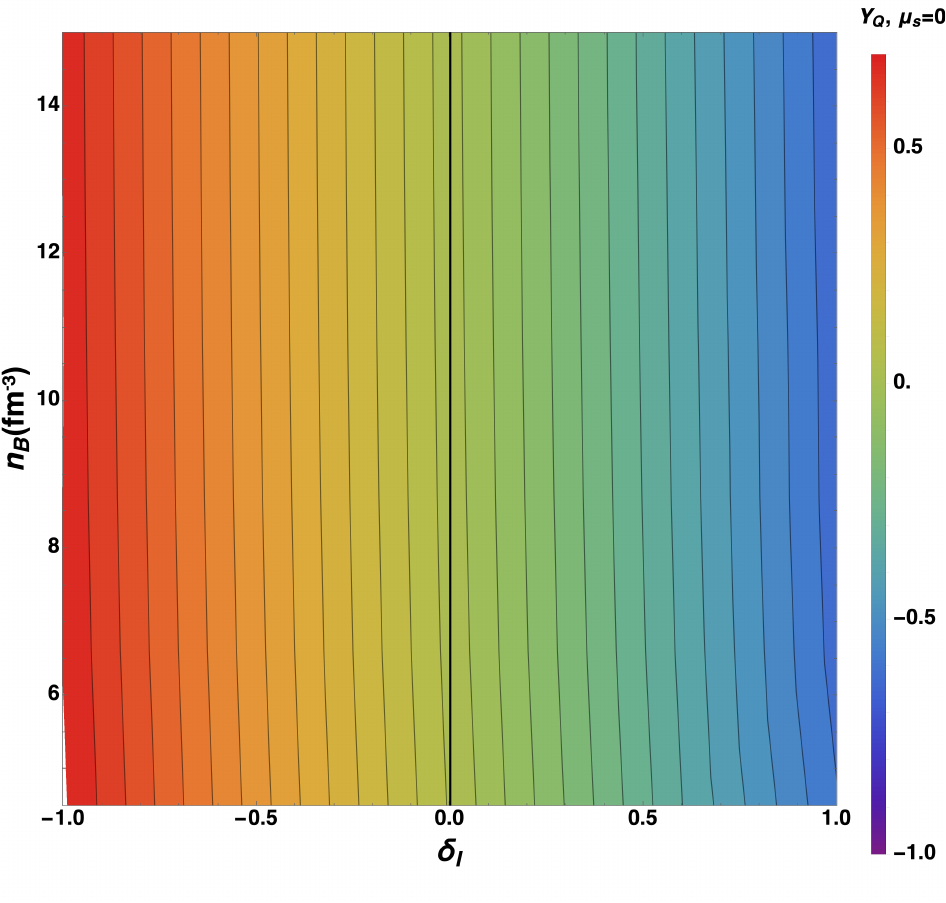}\\
     \end{tabular}
     \caption{
     Density plot of the strangeness (top) and electric charge (bottom) fractions, $Y_S$ and $Y_Q$, as functions of the isospin asymmetry $\delta_I$ and baryon number density $n_B$. 
     The left column shows the limit of isospin symmetric matter where $\mu_S=-1/2\mu_Q$, while the right column shows electroweak equilibrium where $\mu_S=0$. 
     }\label{fig:YS_YQ}
\end{figure*}

Before continuing, it is important to discuss two limits regarding the symmetry energy expansion with strangeness.  
At finite chemical potentials, it is possible to lie on a trajectory with perfect isospin symmetry \cite{Aryal:2021ojz,Yang:2025wop} when the chemical potentials are constrained to be $\mu_S=-1/2\mu_Q$.  In this limit, the isospin symmetry expansion is sufficient up to $\mathcal{O}(\delta_I^2)$ and no skewness terms are required.  The system is isospin symmetric because $Y_S\left(\delta_I\right)$ is also symmetric around the SNM axis of $\delta_I=0$. While we are not aware of any systems that probe this regime of the QCD phase diagram, it can be useful to discuss as a simplification for the strange symmetry energy expansion. 

Results for this isospin symmetric regime with $\mu_S=-1/2\mu_Q$ can be found on the left side of Fig.\ \ref{fig:YS_YQ}. 
The strangeness fraction $Y_S(n_B, \delta_I)$ is shown on the top-left corner, where  it is clear that, as expected, the strangeness fraction is symmetric around the $\delta_I=0$ axis, and thus preserves the isospin reversal symmetry under $\delta_I\to -\delta_I$. 
The charge fraction $Y_Q(n_B, \delta_I)$ is shown on the bottom left, where it is seen to decrease with increasing $\delta_I$.
Note that, for asymptotically large densities and $\delta_I=0$, an equal proportion of the three quark flavors, and thus charge neutrality $Y_Q=0$, is expected. 
Hence, because $Y_S$ is kept close to $-1$ (roughly one third of strange quarks), we find that $Y_Q$ is nearly vanishing at $\delta_I=0$, and approaches $Y_Q(\delta_I=0)\approx 0^+$ as $Y_S$ approaches $-1$ from above at higher densities. 
In fact, $Y_Q(\delta_I)$ slightly decreases with $n_B$ at any fixed value of $\delta_I$. 

Nonetheless, isospin-breaking electroweak equilibrium is the relevant limit for neutron stars that have lived long enough for all their weak interactions to reach chemical equilibrium. In this limit, a few conditions are imposed:  
\begin{enumerate}
    \item  $\mu_S=0$, where this is the \emph{strangeness} chemical potential, not the one for the strange quark, related to the former according to $\mu_s=1/3\mu_B-1/3\mu_Q-\mu_S$. We use $s$ for the strange quark and $S$ for strangeness throughout this paper.
    \item $\mu_Q=-\mu_l$, where $\mu_Q$ is the QCD electric charge chemical potential and $\mu_l$ is the leptonic chemical potential(s).
    \item $n_Q=-\sum_l n_{Q,l}$, that is, to ensure electric neutrality, the net electric charge of QCD particles (nuclei, hadrons, or quarks) must be equal and opposite to the net electric charge of all leptonic particles. 
\end{enumerate}
In this work, we are only considering the QCD contributions such that we only impose condition 1, but not conditions 2--3 (this is left for later work). That is, we work in a regime of (partial) strange electroweak equilibrium, where only the strangeness content is in chemical equilibrium, but not the isospin. 
Results under condition 1 are shown on the right side of Fig.\ \ref{fig:YS_YQ}, where we show 
 $Y_Q(n_B,\delta_I)$ (top) and $Y_S(n_B,\delta_I)$ (bottom) for $\mu_S=0$.

 On the top-right corner of Fig.\ \ref{fig:YS_YQ}, 
 we can see that strange electroweak equilibrium, $\mu_S=0$ explicitly breaks the isospin reversal symmetry.
 That is, the reflection symmetry around $\delta_I=0$ found for isospin-symmetric matter on the top-right corner is completely lost.  
 In fact,  $Y_S|_{n_B}$ significantly decreases with $\delta_I$  for all values of $n_B$, and the range in $Y_S(\delta_I)$ is significantly larger for strange  electroweak equilibrium. 
 In stark contrast to the symmetric case with $\mu_S=-\mu_Q/2$ where $Y_S\approx -1$ within $10\%$, here $Y_S$ can vary by up to a factor of 2, from $-0.6$ to $-1.2$.
 Most strikingly, values of $Y_S<-1$ indicate that strange quarks can even dominate the system for values of $\delta_I\gtrsim 0.5$. 
 Moreover, because strange quarks also contribute to the binding energy, the breaking of isospin reversal symmetry seen in $Y_S(\delta_I)$ leads to a non-zero skewness term $\tilde{E}_{sym,3}(n_B)\neq 0$~\cite{Yang:2025wop}. 
Comparing the skewness in $Y_S(\delta_I)$ to what was previously found for hadronic matter with hyperons in \cite{Yang:2025wop}, we find that pQCD leads to a sharper change in $Y_S$ with $\delta_I$ such that the two are directly related to each other (e.g. large $\delta_I$ gives a large, negative $Y_S$ vs very negative $\delta_I$ that gives a small, negative $Y_S$). 
The difference with respect to hadronic matter can be attributed to the fewer possible BSQ charge combinations that come with 3 flavors of quarks compared to the entire baryon octet and decuplet.

On the bottom-right side of Fig.\ \ref{fig:YS_YQ}, 
$Y_Q(\delta_I,n_B)$ generally shows a similar qualitative behavior to what was seen for isospin symmetric matter, in that $Y_Q(\delta_I)|_{n_B}$ decreases with increasing $\delta_I$. 
However, because $Y_S$ is very sensitive to $\delta_I$, $Y_Q(\delta_I)=(1-\delta_I + Y_S(\delta_I))/2$
varies within a larger range.
That is, in strange electroweak equilibrium, the indirect dependence via $Y_S$ amplifies the dependence  of the charge fraction on $\delta_I$. 
Thus, we expect that matter at electroweak equilibrium has a larger symmetry energy.

 Let us now discuss the possibility of quark matter being charge neutral on its own, that is, $Y_Q=0$. 
For equal quark masses, an equal proportion of flavors $n_u=n_d=n_s$ ($Y_S=-1$, $\delta_I=0$, $Y_Q=0$) is found for $\mu_Q=\mu_S=0$, so that leptons are not required to neutralize the system in electroweak equilibrium. 
In reality, however, the strange quarks are suppressed by their larger mass, leading to $Y_S>-1$, so that $Y_Q=0$ would correspond to $\delta_I>0$ (see Eq.~\eqref{eq:deltaI}), as can be seen upon careful examination of the bottom panels of Fig. \ref{fig:YS_YQ}. 
This, in turn, would require a finite value of $\mu_Q<0$, which is incompatible with the absence of leptons. 
Consequently, in both the isospin-symmetric and electroweak equilibrium cases discussed here, 
leptons are required to achieve electric charge neutrality, due to the strange quark mass $m_s\neq 0$.
At asymptotically high densities,  as the strange quark mass becomes irrelevant compared to $\mu_B$, $Y_S\to -1$ for $\delta_I=0$, as can also be seen from Fig.\ \ref{fig:YS_YQ}. Accordingly, a smaller lepton fraction is required for charge neutrality as $n_B$ increases.

\section{Symmetry Energy Expansion}

\begin{figure}
    \centering
    \includegraphics[width=1.0\linewidth]{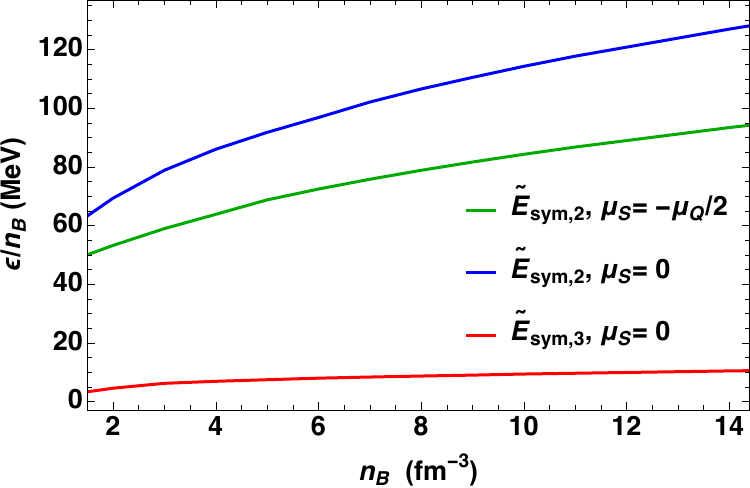}
    \caption{Quadratic and cubic (or skewness) symmetry energy coefficients, $\tilde{E}_{sym,2}$ and $\tilde{E}_{sym,3}$, as functions of the baryon number density, both for isospin-symmetric ($\mu_S=-\mu_Q/2$) and weak-equilibrium ($\mu_S=0$) matter. In the former case, the skewness term vanishes and therefore is not shown. 
    }
    \label{fig:S2_S3}
\end{figure}

Now that we have a better understanding of the phase diagram of cold, dense quark matter from pQCD, we can calculate its symmetry energy in the two limits of isospin symmetric matter and strange electroweak equilibrium.  
First, for isospin symmetric matter, we anticipate to only obtain a quadratic $\tilde{E}_{sym,2}(n_B)$ term in the expansion from Eq.\ (\ref{eqn:symexpan}). 
In the case of electroweak equilibrium, we anticipate a skewness term will appear i.e. $\tilde{E}_{sym,3}(n_B)\neq 0$, but this term will be significantly smaller when compared to $\tilde{E}_{sym,2}(n_B)$.

In Fig.\ \ref{fig:S2_S3} we show the symmetry energy coefficient for $\delta_I^2$ and $\delta_I^3$. 
We find the same qualitative results that were previously found in \cite{Yang:2025wop} i.e. that 1.) isospin symmetric matter still has reflection symmetry across the $\delta_I=0$ axis such that it only requires  $\tilde{E}_{sym,2}(n_B)$; 2.) electroweak equilibrium breaks isospin reflection symmetry due to the skewness in $Y_S(\delta_I)$, such that a $\tilde{E}_{sym,3}(n_B)$ term appears; and 3.) the quadratic coefficient $\tilde{E}_{sym,2}(n_B)$  for strange electroweak equilibrium is always larger than that for isospin symmetric matter. 
Much of these conclusions can be drawn from the phase diagrams shown in the previous section, as was discussed in the text around Fig.\ \ref{fig:YS_YQ}.

How does the symmetry energy of quarks compare to that of hadrons? It is a bit hard to make a complete apples-to-apples comparison since the work of \cite{Yang:2025wop} used a chiral mean field model based on strong-coupling with self-interactions whereas here we are using pQCD, that explicitly relies on the extremely weak coupling regime. However, even with these caveats in mind, we can at least broadly compare our results in these two limits.

In CMF++, $\tilde{E}_{sym,2}(n_B)\sim [50-200]$ MeV for the range shown (in both limits), whereas here we find that the range is closer to $\tilde{E}_{sym,2}(n_B)\sim [50-120]$ MeV for pQCD (at much higher $n_B$). Both approaches find $\tilde{E}_{sym,2}(n_B)\sim [50-200]$ MeV that always increases with $n_B$, but CMF++ shows a significantly larger symmetry energy at the lower range of $n_B$ (as would nearly any hadronic model).  Thus, if one were to smoothly map from some hadronic model into quarks, there must be a non-monotonic dip in $\tilde{E}_{sym,2}(n_B)$  to signify that transition. 

We now also compare the skewness term that appears for electroweak equilibrium.  In CMF++, the term switches sharply from vanishing to a finite value when strange baryons appear. 
Generally, in a strange hadronic phase from CMF++ the $\tilde{E}_{sym,3}(n_B)$ is significantly larger than what we find in pQCD (approximately 6 times larger). 
In pQCD we find that  $\tilde{E}_{sym,3}(n_B)$ is nearly flat with $n_B$ and remains close to around 10 MeV.

\section{Conclusions and Outlook}

In this work, we have performed the first symmetry energy calculations with NLO pQCD with realistic quark masses at vanishing temperatures. To do so, we took special care in correctly defining the isospin symmetry in the presence of strangeness using the Gell-Mann-Nishijima formula. In our approach, we confirmed that a skewness term appears in the strangeness symmetry energy expansion in the limit of weak-$\beta$-equilibrium.  
Another interesting finding was that the pQCD results produce significantly smaller values of $\tilde{E}_{sym,2}(n_B)$ and $\tilde{E}_{sym,3}(n_B)$ compared to hadronic models such that the appearance of weakly interacting quarks would necessitate  non-monotonic features such as a minimum in the symmetry energy.

Given our results, it would be very interesting if realistic quark masses were added to higher-order pQCD calculations such as N2LO and beyond.  Furthermore, it may be possible to develop thermodynamic constraints from these results, similarly to what was done using stability and causality in \cite{Komoltsev:2021jzg}, but we leave this for a future work. Such an approach would need to consider a more complicated stability matrix for multiple conserved charges at $T=0$, see e.g. Appendix E from \cite{Cruz-Camacho:2024odu}.
Finally, another clear extension of this work would be to repeat these calculations in other models that contain a variety of quark phases where $\mu_B,\mu_S,\mu_Q$ can easily be varied independently, like CMF \cite{Dexheimer:2009hi,Cruz-Camacho:2024odu}, NJL \cite{Klevansky:1992qe,Baym:2019iky,Gholami:2024diy}, quarkyonic matter \cite{McLerran:2007qj,Duarte:2020xsp,Moss:2024uam} etc. It would be especially interesting to compare to strongly interacting quark phases, to understand how stronger or weaker interactions affect the quark symmetry energy.

\section*{Acknowledgments} 

We would like to thank Eduardo Fraga for important discussions on the NLO pQCD calculations, Aleksi Kurkela on discussion about the pQCD constraints for neutron stars, and Mateus Reinke Pelicer for discussions on thermodynamic derivatives. 
This research was
partly supported by the US-DOE Nuclear Science Grant No. DE-SC0023861 and by
the National Science Foundation (NSF) within the framework of the MUSES collaboration, under grant number
OAC-2103680. Any opinions, findings, and conclusions
or recommendations expressed in this material are those
of the author(s) and do not necessarily reflect the views of
the National Science Foundation. We also acknowledge
support from the Illinois Campus Cluster, a computing
resource that is operated by the Illinois Campus Cluster Program (ICCP) in conjunction with the National
Center for Supercomputing Applications (NCSA), which
is supported by funds from the University of Illinois at
Urbana-Champaign.
M.H. was supported by the Brazilian National Council for Scientific and Technological Development (CNPq) under process No. 313638/2025-0.

 \appendix
 \section{Renormalization and Error Bars }
\begin{figure*}[t] 
    \centering
    \begin{subfigure}[b]{0.32\textwidth}
        \includegraphics[width=\linewidth]{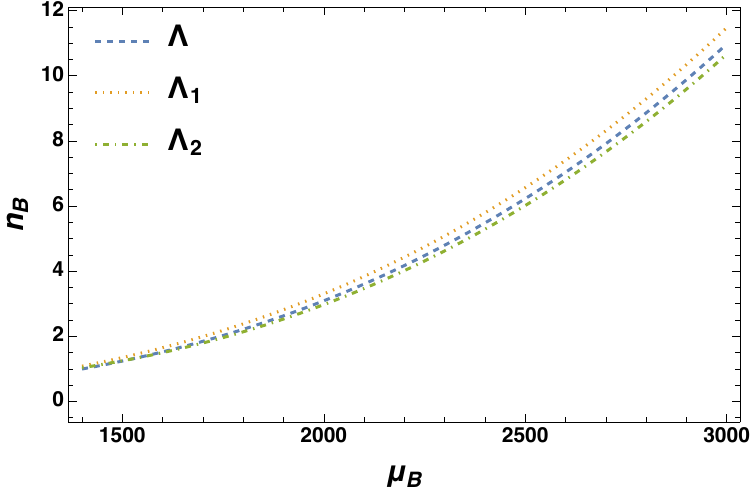}
        \caption{$n_B$}
        \label{fig:error_nb}
    \end{subfigure}\hfill
    \begin{subfigure}[b]{0.32\textwidth}
        \includegraphics[width=\linewidth]{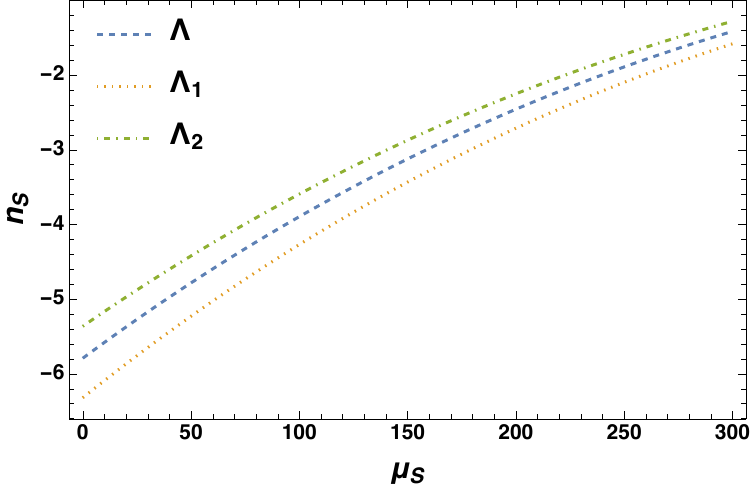}
        \caption{$n_S$}
        \label{fig:error_nS}
    \end{subfigure}\hfill
    \begin{subfigure}[b]{0.32\textwidth}
        \includegraphics[width=\linewidth]{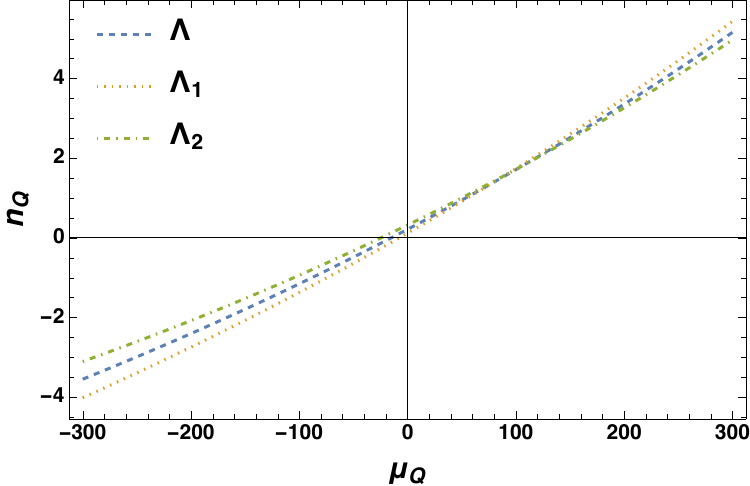}
        \caption{$n_Q$}
        \label{fig:error_nQ}
    \end{subfigure}
    \caption{Effect of changing the renormalization scale according to Eqs.~\ref{Eq:lambda1}–\ref{Eq:lambda2}.}
    \label{fig:errors_combined}
\end{figure*}

 In this Appendix, we estimate the uncertainty in our calculations arising from the choice of renormalization scheme value from Eq~\ref{Eq:lambda}. For that, we define two alternative $\Lambda$'s as:
\begin{equation}
\Lambda_{1}(\mu_{B}, \mu_{Q}, \mu_{S}) = 4\left(\frac{\mu_{B}}{3} + \frac{\mu_{Q}}{6}\right)\, ,
\label{Eq:lambda1}
\end{equation}
\begin{equation}
\Lambda_{2}(\mu_{B}, \mu_{Q}, \mu_{S}) = \frac{3}{2}\left(\frac{\mu_{B}}{3} + \frac{\mu_{Q}}{6}\right)\, .
\label{Eq:lambda2}
\end{equation}
The comparisons are shown in Figs.~\ref{fig:error_nb}–\ref{fig:error_nQ}–\ref{fig:error_nS} for the different densities considered in this work. We have also examined the impact of this variation in Fig.~\ref{fig:S2_S3}. We observe that at lower densities, around $n_B = 2\,\text{fm}^{-3}$, the effect is approximately 5\% for $\mu_S = 0$ and 10\% for $\mu_S = -\mu_Q/2$, which are the regions most sensitive to changes in $\Lambda$. These effects decrease as $n_B$ increases, becoming smaller than 5\% in both cases at $n_B = 15\,\text{fm}^{-3}$.
\bibliography{inspire,NOTinspire}

\end{document}